\documentclass[english,prl,twocolumn,floatfix,showpacs,superscriptaddress]{revtex4}
\usepackage[T1]{fontenc}
\usepackage[latin1]{inputenc}
\usepackage{amsmath}
\usepackage{graphicx}
\usepackage{amssymb}

\makeatletter

\usepackage{times}
\usepackage{amsfonts}
\usepackage{latexsym}

\newtheorem{theorem}{Theorem}

\newenvironment{proof}[1][Proof]{\noindent\textbf{#1.} }{\ \rule{0.5em}{0.5em}}

\usepackage{babel}
\makeatother
\begin{document}

\title{Correlated errors can lead to better performance of quantum codes}

\author{A. Shabani}

\affiliation{Department of Electrical Engineering, University of
Southern California, Los Angeles, California 90089, USA}

\begin{abstract}
A formulation for evaluating the performance of quantum error
correcting codes for a general error model is presented. In this
formulation, the correlation between errors is quantified by a
Hamiltonian description of the noise process. We classify correlated
errors using the system-bath interaction: local versus nonlocal and
two-body versus many-body interactions. In particular, we consider
Calderbank-Shor-Steane codes and observe a better performance in the
presence of correlated errors depending on the timing of the error
recovery. We also find this timing to be an important factor in the
design of a coding system for achieving higher fidelities.
\end{abstract}


\maketitle \subsection{I. INTRODUCTION} The theory of fault-tolerant
quantum computation (FTQC) has been developed as a realistic
extension of the ideal theory of quantum computation. In an ideal
prototype, all computational components are functioning with no
deficiency, while in the real world deviations from the ideal design
is inevitable. This can be due to imperfections in the components or
interactions imposed by the environment. In FTQC, schemes of error
correction are combined with computational elements in a special
design to prevent the spread of errors through a circuit. Applying
error correction steps in a multi-layer circuit is a well
established procedure for FTQC
\cite{Knill:98,Aharonov:99,Kitaev:97,Lidar:06,Aliferis:05,Terhal:05,Aharonov:06}.

Inspired by classical coding theory, a fault-tolerant form of
quantum error correction codes (QECCs) can be achieved by
concatenating single units of encoding. The performance of
fault-tolerant codes has been studied for various noise models. The
classical model of a Markovian-independent noise has been thoroughly
investigated in the literature
\cite{Knill:98,Aharonov:99,Kitaev:97,Lidar:06}. In this picture,
each individual component of the circuit is only coupled to a single
designated bath, with no cross talk between the baths. In addition,
each bath is assumed to be large enough to constitute a local
Markovian decoherence channel. Other possible defects coming from
imperfections in computational gates are also assumed to have local
and instantaneous effects. All these give rise to a probabilistic
description of the noise process. Upper bounding the error
probability guarantees an arbitrary small likelihood of
computational failure.

However, an analysis of quantum computer proposals reveals the
inaccuracy of the above noise model. For example, ion qubits in an
ion trap setup are collectively coupled to their vibrational modes
\cite{Garg:96}. In a quantum dot design, different qubits are
coupled to the same lattice, thus interacting with a common phonon
bath \cite{Loss:98}. The exchange interaction is the main candidate
for implementing two-qubit gates in solid-state proposals. Recently,
it has been shown that many-body exchange interactions can be strong
enough to act as a major source of noise \cite{Mizel:04}. Equally
important, collective control of the qubits may also give rise to
errors due to the inaccuracy of the control field \cite{Wu:04}.
These examples invalidate the assumption of error independence
(locality in space). In addition, the non-Markovian nature of noise
has been observed in various systems \cite{Breuer:Book}. Therefore,
the assumptions of exact locality in time and space should be
relaxed to attain a more realistic model \cite{Gottesman:07}. First
attempts to introduce correlations into noise models were in a
classical manner: multilocation joint probabilities that are
stronger than the independent model \cite{Aharonov:99,Knill:98}.
Recently a physical approach to the problem was taken by introducing
a Hamiltonian description of the noise
process\cite{Aliferis:05,Aharonov:06}. However, because of the
discrete nature of error correction, the noise Hamiltonian was
perturbatively treated to achieve a faulty paths description of\ the
noisy computation. In another recent paper \cite{Gilbert:07},the
authors explicitly show the importance of a non-perturbative
calculation of the FTQC threshold.

In this paper, we study the performance of quantum codes in a single
block of error correction towards a comprehensive analysis of the
FTQC problem with a realistic model. Our aim is to compare spatial
correlations in noise without any limiting assumptions on time.
Focusing on well-defined Calderbank-Shor-Steane (CSS) codes
\cite{CSS}, we first formulate a measure for the code performance
and then apply it to different decoherence scenarios.

Our main finding is that a coding system can function better in the
presence of error correlations. This is mainly due to the entangling
power of a common bath which can help in the preservation of the
code-state coherence designated for recovery \cite{Benatti:05}. This
is in contrast to results from previous studies
\cite{Klesse:05,Novais:06,Duan:99,Clemens:04}.

\subsection{II. CORRELATED ERRORS} First we define the notion of
correlation in errors. From the physical point of view, the primary
probabilistic description of the noise originates from two main
assumptions on the system-bath interaction: each qubit is coupled to
its own bath, and the resulting decoherence is described by a
completely positive (CP) map. A CP-map modeling of the noise process
allows a probabilistic interpretation of the errors which may be
otherwise impossible \cite{Shabani:05}. Based on this classical
randomness picture, correlations could be defined by some joint
probability distribution for the multi-qubit error operators.
However, a microscopic description is lacking in such a model,
motivating a more physical formulation. We remark that the source of
error is either a second system with a large Hilbert
space or an imperfect driving field of a quantum gate. %

\begin{figure}
\includegraphics[width=0.5\textwidth]{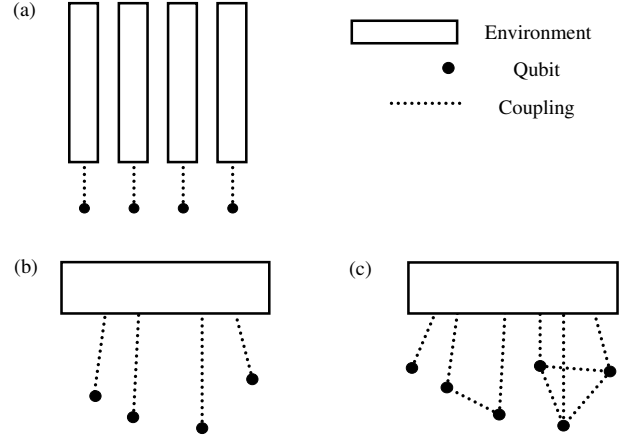} \caption{\label{fig1} A sketch of (a) an uncorrelated error model, (b) a short-range correlation, and (c) a long-range correlation.}
\end{figure}

In general there are three possible scenarios as depicted in Fig.
\ref{fig1}: (i) errors with no correlation, which corresponds to
model of separated baths, each coupled only to a single qubit; (ii)
short-range correlations in which all the qubits are only coupled to
a common bath; and (iii) The most general case with long-range
correlations where qubits are interacting with each other while
sharing a common environment.

In the following, we locate the qubits of a code in the above scenarios
and compare their performance as a function of decoherence time $\tau_{d}$.
This is when we enter the recovery phase. The choice of CSS code is
made to obtain an analytical solution to the problem. A $[n,k,d]$
CSS code is an $n-$qubit code, encoding $k$ bits of information,
capable of correcting errors acting on at most $t=[\frac{d-1}{2}]$
different qubits. Consider an $n$-qubit quantum code in a single
step of error correction. The $n$ qubits are sent through a noise
channel $\mathcal{N}$ expressed as a completely positive (CP) map:
$\mathcal{N(\rho}_{S}\mathcal{)=}\sum$ $_{\alpha}E_{\alpha}\mathcal{\rho}_{S}E_{\alpha}^{\dagger}.$
We expand this map in the basis $\{\Sigma_{\alpha_{1},...,\alpha_{n}}=\sigma_{\alpha_{1}}^{1}\otimes...\otimes\sigma_{\alpha_{n}}^{n}\},$\begin{equation}
\mathcal{N(\rho}_{S}\mathcal{)=}\sum_{\substack{\{ p\}=p_{1},...,p_{n},\\
\{ q\}=q_{1},...,q_{n}} }e_{\{ p\},\{ q\}}\Sigma_{\{ p\}}\text{
}\mathcal{\rho}_{S}\text{ }\Sigma_{\{
q\}}\label{expansion}\end{equation} where $\{\sigma_{0}=I,$
$\sigma_{1}=\sigma_{x},$ $\sigma_{2}=\sigma_{y},$
$\sigma_{3}=\sigma_{z}\}.$ Following the matrix product structure of
the basis, we introduce an $w$th-level submap
$\mathcal{N}_{w}\mathcal{(\rho}_{S}\mathcal{)}$ to be the sum of the
terms in the above expansion (\ref{expansion}) of which the array
$({p_{1}+q_{1},...,p_{n}+q_{n}})$ has $w$ non-zero elements. The
matrix representation of the superoperator $\mathcal{N}_{w}=[e_{\{
p\},\{ q\}}^{w}]$ has the following elements:
\begin{equation}
e_{\{ p\},\{ q\}}^{w}=\left\{ \begin{array}{c}
e_{\{ p\},\{ q\}}\text{ \ \ \ }(p_{i}+q_{i})\text{ has }w\text{ nonzero elements,}\\
0\text{ \ \ \ \ \ \ \ \ \ \ \ \ \ \ \ \ \ \ otherwise. \ \ \ \ \ \ \ \ \ \ \
\ \ \ \ \ \ \ \ \ \ \ \ \ \ \ \ }\end{array}\right.\label{level-n}\end{equation}

The noise map $\mathcal{N(\rho}_{S}\mathcal{)}$ can be decomposed
as a sum of sub-maps $\mathcal{N}_{w}$\begin{equation}
\mathcal{N(\rho}_{S}\mathcal{)=}\sum_{w}\mathcal{N}_{w}\mathcal{(\rho}_{S}\mathcal{)}=\sum_{w,\{ p\},\{ q\}}e_{\{ p\},\{ q\}}^{w}\Sigma_{\{ p\}}\text{ }\mathcal{\rho}_{S}\text{ }\Sigma_{\{ q\}}\label{NL-exp}\end{equation}

We can separate the correctable (by means of a $[n,k,d]$ CSS code)
and uncorrectable parts of the map by splitting the noise map into
submaps $\mathcal{N}_{c}\mathcal{(}\rho_{S}\mathcal{)=}\sum_{w\leqslant t}\mathcal{N}_{w}\mathcal{(}\rho_{S}\mathcal{)}$
and $\mathcal{N}_{ic}\mathcal{(}\rho_{S}\mathcal{)=}\sum_{t<w}\mathcal{N}_{w}\mathcal{(}\rho_{S}\mathcal{)}$:

\begin{equation}
\mathcal{N(}\rho_{S}\mathcal{)=N}_{c}\mathcal{(}\rho_{S}\mathcal{)+N}_{ic}\mathcal{(}\rho_{S}\mathcal{)}.\label{decomp.}\end{equation}

The correctibility of the linear map $\mathcal{N}_{c}$ follows from
the fact that a Hermitian linear map can be corrected by quantum
codes \cite{Shabani:06}. By a Hermitian map we mean a linear map
$\Phi:\mathcal{M}_{u}\longrightarrow\mathcal{M}_{v}$ (where
$\mathcal{M}_{m}$ is the space of $m\times m$ matrices) with a
symmetric representation $\Phi(\rho)$ $=\sum\nolimits
_{\alpha}c_{\alpha}F_{\alpha}\rho F_{\alpha}^{\dagger}$, where the
{}``map operators\textquotedblright\ $\{ F_{\alpha}\}$ are $v\times
u$ matrices and the $c_{\alpha}$s are real numbers.

\begin{theorem} Consider a CP noise map $\Phi^{CP}(\rho)=\sum\nolimits _{\alpha}F_{\alpha}\rho F_{\alpha}^{\dagger}$
and a code space $P_{C}$. A recovery CP map $\mathcal{R}$\ correcting
the map $\Phi^{CP}(\rho)$ can also correct a Hermitian map $\Phi(\rho)=\sum\nolimits _{\alpha}c_{\alpha}F_{\alpha}\rho F_{\alpha}^{\dagger}.$
\label{thrm1} \end{theorem}

\begin{proof} This is a straightforward application of theorem 1
in Ref. \cite{Shabani:06}. \end{proof}

It is obvious that the correctable sum noise map $\mathcal{N}_{c}$
is a Hermitian linear map by definition. Therefore one can find a
recovery map $\mathcal{R(}\rho\mathcal{)}$ inverting
$\mathcal{N}_{c}$, which belongs to the correctibility domain of a
$[n,k,d]$\ code$,$; i.e. there exists a real number
$p_{\mathcal{N}}$ such that\
\begin{equation}
\mathcal{R\circ
N}_{c}=p_{\mathcal{N}}\mathcal{I}.\label{correct}\end{equation}

where $\mathcal{I}$ is the identity superoperator.

\subsection{III. PERFORMANCE MEASURE} We aim to study the performance
of a quantum code for different decoherence processes; therefore, we
need a proper measure to quantify it. As we discussed, the noise map
$\mathcal{N}$ is partially correctable by the recovery map
$\mathcal{R}$. We introduce $p_{\mathcal{N}}$ in Eq. (\ref{correct})
to be the \char`\"{}performance measure\char`\"{} of a code. Before
a rigorous derivation\ of this measure, let us apply it to a model
of independent single-qubit noise channels
\begin{equation}
\mathcal{N}^{indp}\mathcal{(\rho}_{S}\mathcal{)=\{
E}_{1}\otimes...\otimes\mathcal{E}_{n}\}\mathcal{(\rho}_{S}\mathcal{)}.\label{local-noise}\end{equation}
where $\mathcal{E}_{i}(\rho)=(1-p)\rho+p\epsilon(\rho)$, with $p$
the probability of the error channel $\epsilon$. Suppose
$\mathcal{R}$ is the recovery channel correcting errors in up to
$t$\ different locations in the code; then we have\begin{multline}
\mathcal{R\circ N}^{indp}\mathcal{(\rho}_{S}\mathcal{)}=\mathcal{R\circ}\{\mathcal{N}_{c}^{indp}\mathcal{(\rho}_{S}\mathcal{)+N}_{ic}^{indp}\mathcal{(\rho}_{S}\mathcal{)}\}\notag\\
=\sum_{c=0}^{t}(1-p)^{(1-c)}p^{c}\mathcal{R\circ}\sum_{\substack{r_{1}+...+r_{c}\\
=n-c}
}\mathcal{\{}I^{\otimes r_{1}}\otimes\epsilon...\epsilon\otimes I^{\otimes r_{c}}\}\mathcal{(\rho}_{S}\mathcal{)}\notag\\
+\sum_{c=t+1}^{n}(1-p)^{(1-c)}p^{c}\mathcal{R\circ}\sum_{\substack{r_{1}+...+r_{c}\\
=n-c}
}\mathcal{\{}I^{\otimes r_{1}}\otimes\epsilon...\epsilon\otimes I^{\otimes r_{c}}\}\mathcal{(\rho}_{S}\mathcal{)}\\
=\left(\sum_{c=0}^{t}\binom{n}{c}(1-p)^{(1-c)}p^{c}\right)\mathcal{\rho}_{S}\text{
\ }\mathcal{+}\text{ (error).}\end{multline}

By definition of the performance measure, $p_{\mathcal{N}^{indp}}$
is equal to $\sum_{c=0}^{t}\binom{n}{c}(1-p)^{(1-c)}p^{c}.$ This
value can be interpreted in the language of probability as the
likelihood of having fewer than $t$ individual errors
\cite{Nielsen:Book}. In the above definition we discarded
$\mathcal{R\circ N}_{ic}$ as the error. Actually this term vanishes
in an averaging procedure introduced in Ref. \cite{Klesse:05}.
Consider a code word $\rho_{S}=|\Psi\rangle\langle\Psi|$
experiencing the noise $\mathcal{N(}\rho_{S}\mathcal{)}$ and being
corrected by a CSS code recovery map \begin{equation}
\mathcal{R(}\rho\mathcal{)=}\sum\limits _{\substack{\upsilon\in\{0,1\}^{n},\mu\in\{0,3\}^{n}\\
|\upsilon|,|\mu|\leq t}
}P\Sigma_{\{\upsilon\}}\Sigma_{\{\mu\}}\rho\Sigma_{\{\mu\}}\Sigma_{\{\upsilon\}}P.\end{equation}
Our method is similar to the one introduced in a pertinent recent
work \cite{Klesse:05} for CSS codes, but we generalize it to an
arbitrary noise model. We choose fidelity \cite{Nielsen:Book} as a
proper measure of code performance: \begin{equation}
\mathcal{F(}\rho_{S},\mathcal{R(N(}\rho_{S}\mathcal{)))=}\langle\Psi|\mathcal{R(N(}\rho_{S}\mathcal{))}|\Psi\rangle.\end{equation}
 An analytical criterion is achieved by averaging the fidelity over all code words $|\Psi\rangle$. A CSS quantum code $|\mathcal{C}\rangle$
is spanned by the code words $|\mathcal{Q}{\small\rangle}$ where
$\mathcal{Q}$ is a coset of $C_{2}^{\bot}$ and $C_{1}^{\bot}$,
($C_{i}^{\bot}$ is the orthogonal space to $C_{i},$ $C_{2}\subset
C_{1}\subset\mathbb{Z}_{2}^{n}$). Applying the average over all
pairs $C_{2}\subset C_{1}$ suppresses all terms with $e_{\{ p\},\{
p\}}^{n>d}$ or $e_{\{ p\}\neq\{ q\}}^{n}$ which are of the order of
$2^{-\mathcal{O}(n)}.$ We have verified this fact for a general
noise model, justifying the bipartitioning of the noise map in Eq.
(\ref{decomp.}). As a consequence, the fidelity average converges to
a simple form of\begin{equation}
p_{\mathcal{N}}=\mathcal{\mathcal{F}}_{ave}\mathcal{\mathcal{(}}|\mathcal{C\rangle\langle\mathcal{C}|},\mathcal{\mathcal{R(N(}}|\mathcal{C}\rangle\langle\mathcal{C}|\mathcal{\mathcal{)))=}}\sum_{n\leqslant
t,\{ p\}}e_{\{ p\},\{ p\}}^{n}.\label{ave}\end{equation}

Such a code-independent measure equips us with a quantitative tool
to compare different noise models. \bigskip Suppose system and bath
start from an initial product state $\rho
_{SB}(0)=\rho _{S}(0)\otimes \rho _{B}(0),$ with the bath density matrix $%
\rho _{B}(0)=\sum_{i }\lambda_{i} |b_{i }\rangle \langle b_{i }|$.
For a system-bath propagator $U_{\tau _{d}}$ ($\tau _{d}$ is
decoherence time before recovery operation), the dynamical Kraus operators $%
\{E_{i ,j }\mathcal{\}}$ can be written as $E_{i ,j }=\sqrt{%
\lambda_{i} }\langle b_{j }|U_{\tau _{d}}|b_{i }\rangle .$ Now we
can find a closed form for $p_{\mathcal{N}}$ as a function of
$U_{\tau _{d}}$
and $\rho _{S}(0)$:%
\begin{eqnarray*}
p_{\mathcal{N}}(\tau _{d}) &=&\sum_{n\leqslant t,\{p\}}e_{\{p\},\{p\}}^{n}=%
\frac{1}{2^{2n}}\sum_{|\upsilon |\leq t}|\mathrm{Tr}(\Sigma
_{\{\upsilon
\}}E_{i ,j })|^{2} \\
&=&\frac{1}{2^{2n}}\sum_{|\upsilon |\leq t}\lambda_{i}
|\mathrm{Tr}(\Sigma _{\{\upsilon \}}\langle b_{j }|U_{\tau
_{d}}|b_{i }\rangle )|^{2}\implies
\end{eqnarray*}

 \begin{equation}
p_{\mathcal{N}}(\tau_{d})=\frac{1}{2^{2n}}\sum\limits
_{|\upsilon|\leq
t}\mathrm{Tr}[\mathrm{Tr}_{S}(U_{\tau_{d}}^{\dag}\Sigma_{\{\upsilon\}})\mathrm{Tr}_{S}(\Sigma_{\{\upsilon\}}U_{\tau_{d}})\rho_{B}(0)].\label{performance}\end{equation}

\begin{figure}
\includegraphics[width=0.5\textwidth]{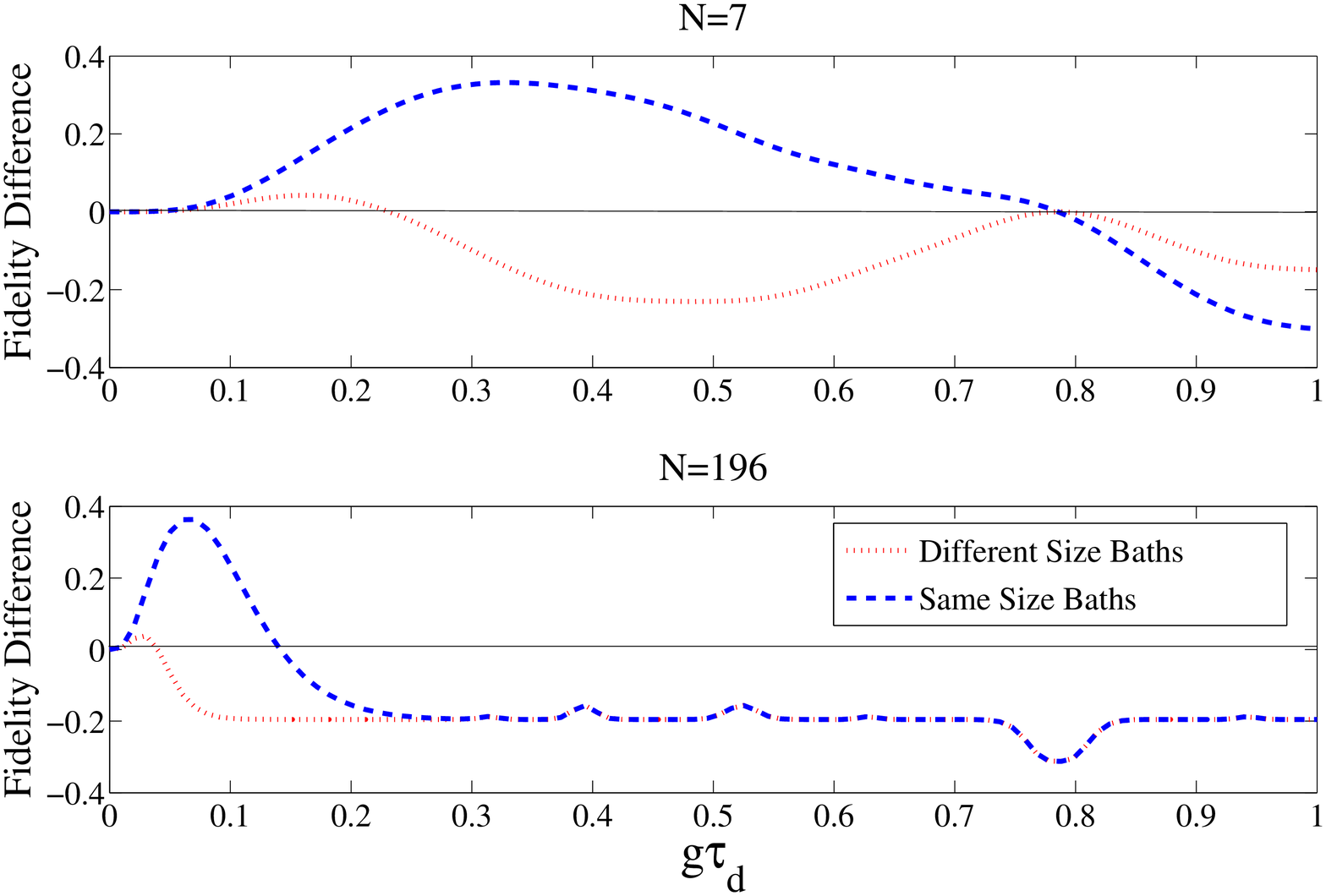} \caption{\label{LvsNL} Performance difference of a [7,1,3] CSS code for two scenarios of local and non-local environments as a function of decoherence time $\tau_{d}$. A negative
     sign $p_{\mathcal{N}}^{local}(g\tau_{d})-p_{\mathcal{N}}^{nonlocal}(g\tau_{d})$ demonstrates better performance for the nonlocal error.($N=\{7,196\}$, $\beta\Omega=0.01$)}
\end{figure}

In the following we apply this measure on different system-bath interaction
configurations.

\subsection{IV. LOCAL vs NONLOCAL ENVIRONMENTS} Here we compare
system-bath configurations (a) and (b) in Fig. \ref{fig1}: Each
qubit is coupled to its own bath versus all qubits coupled to the
same bath. We consider a spin star model which consists of a\
central spin-$\frac{1}{2}$ particle representing a qubit, surrounded
by $N$ localized spin-$\frac{1}{2}$ particles acting as a spin bath
for the qubit \cite{Breuer:04}. In the local model the central spin
$\sigma$ is coupled to the bath spins $\sigma_{i},$ $(i=1,...,N)$
via a dephasing interaction of the form \begin{equation}
H_{hf}=\sum_{i=1}^{N}g_{i}\sigma_{z}Z_{i},\end{equation}
 where $\sigma_{z}$ and $Z_{i}$ are the $z$ components of $\sigma$ and
$\sigma_{i}$ respectively. This is the effective Hamiltonian of
dephasing in NMR due to spin-spin interactions \cite{Ernst:Book}.
Also this hyperfine contact type of interaction is a source of
electron spin decoherence in quantum dots due to the interaction
with nuclei \cite{Khaetskii:02}. In the presence of a strong
magnetic field along the $z$ direction, the dephasing interaction is
the dominant term in the hyperfine interaction \cite{Sousa:03}.

Two different configurations of local and nonlocal decoherence for
an $n$-qubit code are modeled as
$H_{local}=\sum_{m=1}^{n}\sum_{i=1}^{N}g_{im}\sigma_{z}^{m}Z_{i}^{m}$
and
$H_{nonlocal}=\sum_{m=1}^{n}\sum_{i=1}^{N}g_{im}\sigma_{z}^{m}Z_{i}$.
As an example we consider a $[7,1,3]$ code surrounded by
$N=\{7,196\}$ bath spins \cite{Breuer:04}. The bath state,
$\rho_{B}$ in Eq.(\ref{performance}), is a thermal state of the spin
bath at temperature $T=1mK$ with an internal Hamiltonian
$H_{B}=\sum_{i=1}^{N}\Omega Z_{i}.$ In addition, for simplicity we
assume a symmetric coupling $g_{im}=g$. The difference of the
performance of local and nonlocal models
$p_{\mathcal{N}}^{local}(g\tau_{d})-p_{\mathcal{N}}^{nonlocal}(g\tau_{d})$
is plotted in Fig. \ref{LvsNL}. We observe that the performance
strongly depends on the decoherence time $\tau_{d}$. Consequently,
choosing the correct time for recovery can improve the functioning
of the code in the presence of correlations. This is in contrast to
a classical system in which a common source of error acts
simultaneously on different parts, and therefore leads to a higher
risk of computational failure, while in the quantum case, indirect
coupling of the qubits through a shared common bath or direct
coupling by means of many-body interactions imposes additional
dynamics on the code coherence that changes the performance of the
code. Furthermore, these results reveal that engineering the qubits
of a code block in the same location (non-local bath) may have an
advantage over spatially separating them (local bath), as suggested
in \cite{Aharonov:06}.

Another interesting comparison can be made between local and
nonlocal baths of the same size, with the corresponding Hamiltonians
$H_{local}=\sum_{m=1}^{n}\sum_{i=1}^{N/n}g_{im}\sigma_{z}^{m}Z_{i}^{m}$
and
$H_{nonlocal}=\sum_{m=1}^{n}\sum_{i=1}^{N}g_{im}\sigma_{z}^{m}Z_{i}$.
The simulation results are plotted in Fig. \ref{LvsNL} which
demonstrates the time-dependent, alternative behavior of the code
performance difference.

\subsection{V. TWO-BODY vs MANY-BODY INTERACTIONS} 
\begin{figure}
\includegraphics[width=0.5\textwidth]{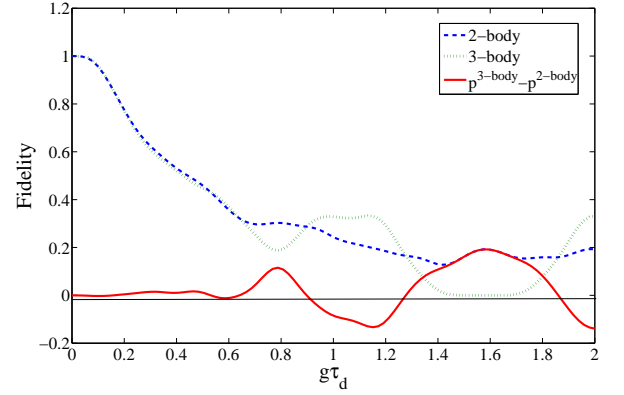} \caption{\label{MB} A [7,1,3] code performance in the presence of two-body and three-body interactions. The performance difference
     $p_{\mathcal{N}}^{two-body}(\tau_{d})-p_{\mathcal{N}}^{three-body}(\tau_{d})$ determines how the effect of three-body interaction varies by decoherence time $\tau_{d}$ ($N=7$).}
\end{figure}

We now introduce
many-body interactions as a counterpart to multivariate joint
probability in classical descriptions of correlations. In a recent
study \cite{Aharonov:06}, the fault-tolerant quantum computation in
the presence of many body interactions has been investigated.
Following their formulation, we include additional terms $H_{ij}$\
which act simultaneously on qubit pairs $<ij>$ and on the bath. As
an example consider the noise Hamiltonian

\begin{equation}
H_{three-body}=g\sum_{j=1}^{n}\sum_{i=1}^{N}\sigma_{z}^{j}Z_{i}+g^{\prime}\sum_{j,k=1}^{n}\sum_{i=1}^{N}\sigma_{z}^{j}\sigma_{z}^{k}Z_{i},\end{equation}

in which $g$ and $g^{\prime}$ determines the strength of two-body
and three-body interactions. The ratio $g^{\prime}/g$, in some
specific configurations of spin-$\frac{1}{2}$ particles, can rise up
to $16\%$ \cite{Mizel:04}. We compare two cases of two-body only
interactions, with $g'=0$, and three-body interactions with
$g^{\prime}=0.1g$. As shown in Fig. \ref{MB}, for short recovery
times they behave similarly, while at longer times an alternative
performance of two cases emerges.

\subsection{VI. FAULTY GATE} A faulty gate is another main source of
errors in computation. In many proposals for quantum computers, a
magnetic field or laser pulse is an ingredient of the control field
which can act locally \cite{Nielsen:Book} or globally (nonlocally)
\cite{Wu:04}. Obviously an inaccuracy in manipulating the control
field causes inaccuracy in the computation. Here we compare the
fault tolerability of an $n$-qubit code driven by an imperfect
magnetic field acting locally $(H_{ml}=g\sum_{i=1}^{n}B_{i}Z_{i})$
or globally $(H_{mg}=gB_{0}\sum_{i=1}^{n}Z_{i})$. Because of the
classical nature of the error, we assume a random process for the
control error, i.e. $B_{i}=B_{ideal}+W_{i}$, where the random
variable$\ W_{i}$ has a probability distribution
$p_{W_{i}}(w)=p(w)$. For a time $\tau_{r}$\ the total rotation
generated by this field is
$U_{ml}(w,\tau_{r})=\exp[-i\tau_{r}g\sum_{i=1}^{n}B_{i}(w)Z_{i}]$
for local control, and
$U_{mg}(w,\tau_{r})=\exp[-i\tau_{r}gB_{0}(w)\sum_{i=1}^{n}Z_{i}]$
for global control. The distance of two unitary operations $U$ and
$V$, acting on a Hilbert space ${\cal {H}}$, can be measured by the
average fidelity Ave$_{|\psi\rangle\in{\cal
{H}}}\langle\psi|V^{\dagger}U|\psi\rangle=\frac{1}{d^{2}}|Tr(V^{\dagger}U)|^{2}$,
with $d=\dim({\cal {H})}$. To quantitatively compare a noisy gate
with the ideal one, we use this distance averaged over the random
variable $W_{i}$. For local control this value becomes
\begin{equation} {\cal
_{F}}_{local}(\tau_{r}g)=(\int_{-\infty}^{\infty}\cos(\tau_{r}g\omega)p(\omega)d\omega)^{n},\end{equation}
 while for the global one it is \begin{equation}
{\cal
_{F}}_{global}(\tau_{r}g)=\int_{-\infty}^{\infty}(\cos(\tau_{r}g\omega))^{n}p(\omega)d\omega.\end{equation}
 Applying the Hölder inequality \cite{Holder}, we find that \begin{equation}
{\cal _{F}}_{local}(\tau_{r}g)\leq{\cal
_{F}}_{global}(\tau_{r}g).\end{equation}
 This indicates the higher fidelity of global control in comparison
to the local one.

\subsection{VII. CONCLUSION}In summary, we have studied quantum
error correction in the presence of correlated errors. A microscopic
description of the noise process enables us to classify correlations
based on their physical relevance in quantum computer designs.
Roughly speaking, we find that it is not necessarily true that
correlations in errors imply additional flaws in computation, but on
the contrary, these correlations may positively affect the
performance. Furthermore, the results presented in this paper
emphasize that, besides choosing a proper code, the pre-recovery
decoherence time is a main factor to achieving higher fidelity,
since the code performance can drastically vary with the timing of
the recovery operation.

\subsection{ACKNOWLEDGMENTS}
The author acknowledges helpful discussions with K. Khodjasteh, D.
A. Lidar, and J. Geraci.

\end{document}